\documentclass{aa}

\usepackage{graphicx}
\usepackage{txfonts}
\usepackage{xcolor}

\begin{document}

\title{Sensitivity of the redshifted 21 cm signal from the Dark Ages \\ to parameters of primordial magnetic fields}

\subtitle{}
\author{B. Novosyadlyj\inst{1}, Yu. Kulinich\inst{1}, N. Fortuna\inst{1}, A. Rudakovskyi\inst{2,3}}

\institute{Astronomical Observatory of Ivan Franko National University of Lviv, Kyryla i Methodia str., 8, Lviv, 79005, Ukraine,
\and Bogolyubov Institute for Theoretical Physics of the NAS of Ukraine, Metrolohichna Str. 14-b, Kyiv 03143, Ukraine
\and Dipartimento di Fisica e Astronomia, Universit\'{a} di Bologna, Via Gobetti 93/2, 40122, Bologna, Italy}

\authorrunning{B. Novosyadlyj et al.}
\titlerunning{21 cm Signal and Primordial Magnetic Fields}

\date{\today}

\abstract
{We analyse the impact of the decaying magnetic turbulence of primordial magnetic fields (PMFs) and ambipolar diffusion on the ionisation and thermal history of the Dark Ages Universe ($30\le z\le300$), and its imprint on the spectral profile of the global signal in the redshifted 21 cm hydrogen line. The heating function caused by decaying magnetic turbulence monotonically decreases after cosmological recombination; its amplitude strongly depends on the strength of the PMFs' $B_0$ and weakly depends on the spectral index of the initial power spectrum of the PMFs' $n_{\rm B}$. The heating function caused by ambipolar diffusion, in contrast, noticeably depends on the spectral index in the range $-3\lesssim n_{\rm B}\lesssim4$ but is subdominant in the Dark Ages epoch for PMF models with $B_0\lesssim0.5$ nG. We computed the ionisation and thermal history of intergalactic gas from the cosmological recombination up to the end of the Dark Ages epoch for a range of PMF parameters, $0.05\lesssim B_0\lesssim0.5$ nG, $-2.9\lesssim n_{\rm B}\lesssim4$, and show the essentially distinguished thermal evolution from one in the Lambda cold dark matter ($\Lambda$CDM) model. We also show that the profile of the redshifted 21 cm hydrogen line is very sensitive to the PMF parameters from this range and can be used for their constraints in conjunction with other observational data.
}
 

\maketitle
%
\section{Introduction} \label{sec:intro}

There is a broad range of observational evidence indicating the existence of magnetic fields in the Universe. Relatively strong fields residing in star systems and galaxies are believed to result from the amplification of seed fields with understood hypothetical magnetodynamos. In contrast, magnetic fields in cosmic voids have no apparent connection to astrophysical sources and support the idea of their cosmological origin (see reviews by \cite{Durrer2013,Subramanian2016}). They are called primordial magnetic fields (PMFs), and the experimental evidence for their existence follows from the observations of blazar radiation in GeV and TeV energy ranges by advanced gamma-ray telescopes \citep{Neronov2010,Dolag2011,Takahashi2013,Tiede2020,Neronov2022}. They give a lower limit on the value of such fields on scales of several megaparsecs: $B_0>10^{-20}$~G. There are a few methods of studying the magnetic fields at cosmological space-time scales, including Faraday rotation measurements \citep{Sullivan_2020,Vazza2024}, magnetic fields' impact on CMB \citep{Paoletti2022}, Ly-$\alpha$ forest method \citep{Pavicevic2025}, dwarf galaxies studies \citep{Sanati2024}, studying high-$z$ structures \citep{Ralegankar2024}, and line-intensity mapping \citep{Adi2023}. The upper limit follows from the study of the cosmic microwave background (CMB) anisotropy by the Planck Space Observatory \citep{Planck2016}: $B_0<9\cdot10^{-10}$~G.

The signals from the Dark Ages, when physical processes in the baryonic component with primordial chemistry were simplest, promise to unveil some puzzles of current cosmology related to primordial fields, the nature of dark matter particles, and others. Such signals are the 21 cm line of atomic hydrogen (see reviews \cite{Barkana2001,Fan2006,Furlanetto2006,Bromm2011,Pritchard2012,Natarajan2014,Shimabukuro2022,Minoda2023}) and the ro-vibrational lines of the first molecules \citep{Kamaya2002,Ripamonti2002,Kamaya2003,Omukai2003,Mizusawa2004,Mizusawa2005,Liu2019,Novosyadlyj2020,Novosyadlyj2022,Maio2022}. They manifest themselves as a weak distortion of the CMB spectrum. In this paper, we discuss the redshifted 21 cm signal in the decametre wavelength range.  

The development of theoretical aspects regarding the dependence of the 21 cm signal from the early Universe on PMFs can be followed in papers \citep{Sethi2005,Tashiro2006,Schleicher2009,Kunze2014,Chluba2015,Kunze2019,Cruz2024,Bhaumik_2025}. They are mainly focused on the global signal at 50-100 MHz expected from the Cosmic Dawn epoch as the strongest one capable of being detected by current advanced antennas,  such as EDGES \citep{Bowman2018} and SARAS~3 \citep{Singh2022}, or power spectrum by array telescopes such as LOFAR\footnote{www.lofar.org}, MWA\footnote{https://www.mwatelescope.ogr}, PAPER\footnote{eor.berkeley.edu/}, GMRT\footnote{http://www.gmrt.ncra.tifr.res.in/gmrt/gmrt.html} or HERA\footnote{ https://reionisation.org/} . 
 
In this paper we study the impact of primordial magnetic fields on the predicted 21 cm hydrogen line signal from the Dark Ages redshifted to the decametre range of electromagnetic waves 10-50 MHz.  
Due to magnetic field decay, there is additional heating of intergalactic gas. The goal is to establish the sensitivity range of this signal to the large-scale averaged amplitude and power spectrum index of PMFs. In previous papers \citep{Novosyadlyj2023,Novosyadlyj2024}, we analysed the sensitivity of the Dark Ages global signal to the strength of nearly scale-invariant PMFs ($n_{\rm B}=-2.9$). In the paper by \cite{Konovalenko2024}, it was shown that the sensitivities of ground-based Ukrainian array telescopes, UTR-2 and GURT, are enough to detect the redshifted 21 signal from the Dark Ages with integration time $\sim60$ days. The most promising ground-based array telescope for investigating the metre-decametre wavelength sky with the highest sensitivity and angular resolution is Square Kilometre Array\footnote{https://www.skao.int/} (SKA), which will start  observations in the coming years. Moreover, a few projects of far-side lunar telescopes aiming to detect this line are under development \citep{Burns2011,Shkuratov2019,Bentum2020,Burns2021,Goel2022,Bale2023}. The Chinese low-frequency radio spectrometer\footnote{https://www.nssdc.ac.cn/nssdc\_en/html/task/change4.html} Chang'E-4 is already in operation and is taking measurements. So detailed investigations of the 21 cm line from the Dark Ages, its dependence on parameters of cosmological models, and the existence of degeneracies between them are important tasks of current cosmology.      

The plan of the paper is as follows. In Sect. II, we describe the state of the intergalactic medium during the Dark Ages, from cosmological recombination up to the beginning of intensive reionisation at $z=10$, in the $\Lambda$CDM cosmological models with different values of amplitude and power spectrum index of PMFs. In Sect. III, we analyse the dependence of the position and profile of the 21 cm absorption line in the Dark Ages on cosmological parameters in the standard $\Lambda$CDM model and additional heating  of the baryonic component in models with PMFs. The comparison of our results with analogical ones of other authors, discussions of disagreements and degeneracies are presented in Sect. IV. The results are summarised in the Conclusions. 
 
\section{Thermal history of the intergalactic medium in the Dark Ages} 

\begin{figure}[htb]
\includegraphics[width=0.49\textwidth]{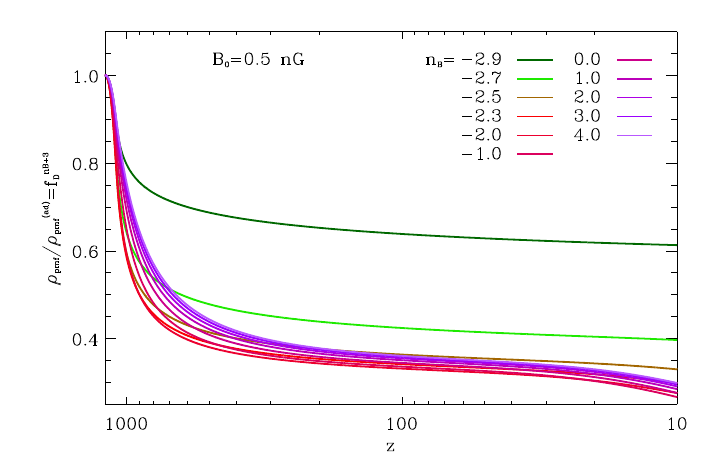}
\caption{Time evolution of PMFs' cut-off scale caused by  decaying magnetic turbulence and ambipolar diffusion ($f_{\rm D}^{n_{\rm B}+3}=\rho_{\rm mf}/\rho_{\rm mf}^{ad}$) for different values of $n_{\rm B}$ (colour lines).}
\label{fD}
\end{figure}

\begin{figure*}[htb]
\includegraphics[width=0.49\textwidth]{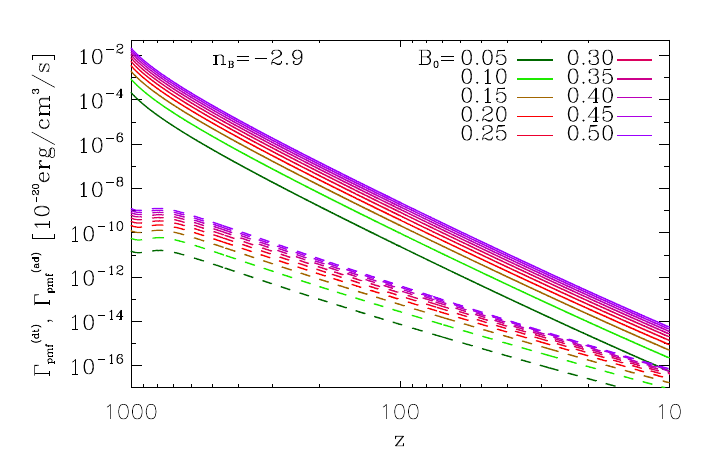}
\includegraphics[width=0.49\textwidth]{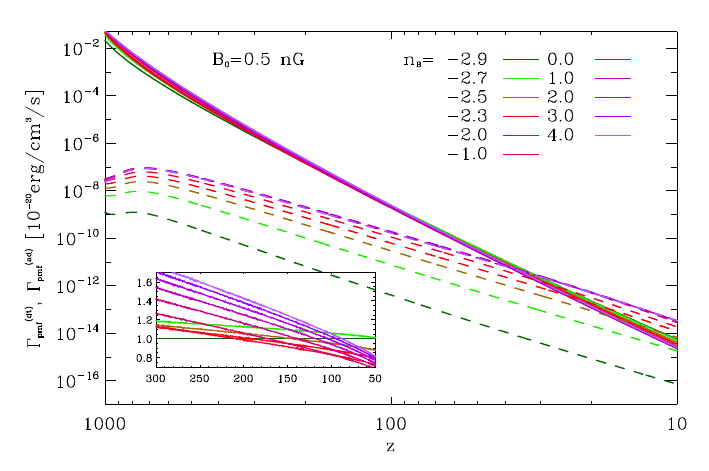}
\caption{Heating functions caused by decaying magnetic turbulence (solid lines) and ambipolar diffusion (dashed lines) in PMFs models with different $B_0$ (left) and $n_{\rm B}$ (right). The inset panel shows $\Gamma_{\rm pmf}^{\rm (dt)}(z,B_0,n_{\rm B})/\Gamma_{\rm pmf}^{\rm (dt)}(z,B_0,n_{\rm B}=-2.9)$ for the purpose of distinguishing lines in the 21 cm line formation region.}
\label{Qpmf}
\end{figure*}

For the calculation of distortion of the CMB thermal spectrum at the redshifted 21 cm line in the Dark Ages epoch, it is necessary to compute the ionisation and thermal history of the gas, as well as the populations of levels of the hyperfine structure of atomic hydrogen. They depend on the parameters of the cosmological model, the ionisation mechanisms, heating, and cooling of the baryonic gas. In this section, we consider the ionisation and thermal history of the gas in the model with primordial magnetic fields and in the standard $\Lambda$CDM model for comparison. In all computations, we assume the cosmological parameters following from \cite{Planck2020a,Planck2020b}: $H_0=67.36\pm0.54$ km/s$\cdot$Mpc (Hubble constant), $\Omega_\mathrm{b}=0.0493\pm0.00112$ (density parameter\footnote{Density parameter is the mean density of the component in the current epoch in units of critical density: $\Omega_i\equiv\rho^0_i/\rho^0_\mathrm{cr}$ where $\rho^0_\mathrm{cr}\equiv 3H_0^2/8\pi G$.} of baryonic matter), $\Omega_\mathrm{dm}=0.266\pm0.0084$ (density parameter of dark matter), $\Omega_r=2.49\cdot10^{-5}$ (density parameter of relativistic component (CMB and neutrinos)), $\Omega_\mathrm{\Lambda}=0.6847\pm0.0073$, and $\Omega_K=0$ (dimensionless\footnote{$\Omega_\mathrm{\Lambda}\equiv c^2\Lambda/3H^2_0$, $\Omega_\mathrm{K}\equiv -c^2K/H^2_0$.} cosmological constant $\Lambda$ and curvature of 3-space $K$, respectively). We assume the primordial chemical composition with the mass fraction of helium $Y_p=0.2446$ \citep{Peimbert2016}, following from the cosmological nucleosynthesis of the standard Big Bang model.

After the epoch of cosmological recombination and before reionisation ($800\lesssim z\lesssim10$), more than  $99$\% of hydrogen was in the neutral state. The residual ionisation of the Universe is a consequence of its expansion. The fraction of ionised hydrogen $x_{\rm HII}$ is less than 1\% at these redshifts and depends on the density, the mechanisms of heating and cooling of the gas, and the expansion rate of the Universe. For a homogeneous expanding Universe, the gas temperature and ionisation of atoms are calculated by integrating a system of differential equations describing energy balance and ionisation-recombination kinetics of atoms\footnote{The first molecules make up a small fraction of all particles (see \cite{Novosyadlyj2020,Novosyadlyj2022}) and are not discussed here in the context of ionisation history.}. In the standard $\Lambda$CDM  model, the main mechanism of gas heating in this period is Compton scattering of CMB radiation on free electrons and adiabatic cooling caused by cosmological expansion. All other heating and cooling mechanisms that occur in astrophysical plasma under Dark Ages conditions are significantly weaker (see Fig. 4 in \cite{Novosyadlyj2023}). 

In our computations, during cosmological recombination and in the epoch of the Dark Ages, the temperature of the plasma and the ionisation of hydrogen and helium up to $z=200$ are calculated using an effective three-level atomic model implemented in the publicly available code RecFast\footnote{http://www.astro.ubc.ca/people/scott/recfast.html} \citep{Seager1999,Seager2000}. The calculation for $z<200$ is by the numerical integration of the system of differential equations:
\begin{eqnarray}
&&\hskip-0.5cm -\frac{3}{2}n_{\rm tot}k_{\rm B}(1+z)H\frac{dT_{\rm b}}{dz} = \Gamma_{\rm CMB}-\Lambda_{\rm ad}- \Lambda_{\rm ff,phr,H_2} + \Gamma_{\rm pmf},\label{Tbe}\\
&&\hskip-0.5cm -(1+z)H\frac{dx_{\rm HII}}{dz}=R_{\rm HI}x_{\rm HI}+C^i_{\rm HI} n_i x_{\rm HI}-\alpha_{\rm HII}x_{\rm HII}x_{\rm e}n_{\rm H}, \label{kes}\\
&&\hskip-0.5cm -(1+z)H\frac{dx_{\rm HeII}}{dz}=R_{\rm HeI}x_{\rm HeI}+C^i_{\rm HeI} n_i x_{\rm HeI}-\alpha_{\rm HeII}x_{\rm HeII}x_{\rm e}n_{\rm H}, \label{kes2}\\
&&\hskip-0.5cm H=H_0\sqrt{\Omega_{\rm m}(1+z)^3+\Omega_{\rm K}(1+z)^2+\Omega_{\Lambda}}, \label{H}
\end{eqnarray}
where $T_{\rm b}$ is the temperature of the baryonic gas; $n_{\rm tot}\equiv n_{\rm HI}+n_{\rm HII}+n_{\rm e}+n_{\rm HeI}+n_{\rm HeII}+n_{\rm HeIII}$ is the number density of particles of all types; $n_{\rm H}\equiv n_{\rm HI}+n_{\rm HII}$ is the number density of hydrogen; $n_{\rm e}\equiv n_{\rm HII}+n_{\rm HeII}+2n_{\rm HeIII}$ is the number density of electrons; $x_{\rm e}\equiv n_{\rm e}/n_{\rm H}$, $x_{\rm HI} \equiv n_{\rm HI}/n_{\rm H}$, $x_{\rm HII} \equiv n_{\rm HII}/n_{\rm H}$, $x_{\rm HeI} \equiv n_{\rm HeI}/n_{\rm He}$, $x_{\rm HeII} \equiv n_{\rm HeII}/n_{\rm He}$, $k_{\rm B}$ is the Boltzmann constant; $\Gamma_{\rm CMB}$ is the heating function due to Compton scattering of CMB quanta on free electrons \citep{Seager1999,Seager2000};    
$\Lambda_{\rm ad}$ is the adiabatic cooling function due to the expansion of the Universe \citep{Seager1999}; $\Lambda_{\rm ff}$ is the cooling function due to free-free transitions of electrons \citep{Shapiro1987}; $\Lambda_{\rm phr}$ is the cooling function due to photorecombination \citep{Anninos1997}; $\Lambda_{\rm H_2}$ is the cooling function due to emission in rotational-vibrational transitions of the $H_2$ molecule \citep{Seager1999}; $\Gamma_{\rm pmf}$ is the heating function caused by PMFs decaying; $R_{\rm HI},\,R_{\rm HeI}$ are the photoionisation rates of hydrogen and helium atoms; $C^i_{\rm HI},\,C^i_{\rm HeI}$ are the collisional ionisation coefficients of the $i$-th type of particles; $\alpha_{\rm HII},\,\alpha_{\rm HeII}$ are the recombination coefficients of hydrogen and helium ions; and $\Omega_{\rm m}=\Omega_{\rm b}+\Omega_{\rm dm}$, $H(z)$ is the expansion rate of the Universe. The reaction rates used here can be found in Appendix of \cite{Novosyadlyj2023}. 
  
We assume that the non-helical magnetic field in the Dark Ages is an isotropic and homogeneous random process, and the following equations by \cite{Mack2002,Sethi2005,Kunze2014,Chluba2015,Minoda2019} describe its statistical properties in terms of a power spectrum, 
\begin{equation}
\langle \tilde{B}^*_i(\mathbf{k})\tilde{B}_j(\mathbf{k'})\rangle=4\pi^3\delta_{\rm D}(\mathbf{k}-\mathbf{k'})(\delta_{ij}-k_ik_j)P_{\rm B}(k),
\end{equation}
where $\tilde{B}_i(\mathbf{k})$ is a Fourier component of $B_i(\mathbf{x})$ with wave vector $\mathbf{k}$ and 
\begin{equation}
 P_{\rm B}(k)=\frac{(2\pi)^{n_{\rm B}+5}}{\Gamma(\frac{n_{\rm B}+3}{2})}\langle B_\lambda^2\rangle\frac{k^{n_{\rm B}}}{k_{\lambda}^{n_{\rm B}+3}},  
 \label{Pk}
\end{equation}
where $n_{\rm B}$ is the spectral index, $\tilde{B}_\lambda$ is the Fourier mode of the smoothed field $B_\lambda$ over scale $\lambda$. Magnetohydrodynamics of the baryon-photon plasma before recombination predicts the decay of magnetic fields at small scales. The smallest scale that survives is defined by the field strength, the dimensionless Alfvén velocity, and the baryon-photon Silk damping scale \citep{Subramania1998}. The damping scale at the recombination epoch is as follows:
\begin{eqnarray}
\hskip-0.5cm k_{\rm D}^{rec}&=&\left[7.32\cdot10^4\left(\frac{k_\lambda}{{\rm Mpc}^{-1}}\right)^{n_{\rm B}+3}\left(\frac{\omega_{\rm m}}{0.15}\right)^{\frac{1}{2}}\left(\frac{\omega_{\rm b}}{0.02}\right)\right.  \\
\hskip0.5cm&\times&\left(\frac{x_{\rm e}^{rec}}{0.1}\right)\left.\left(\frac{1+z_{\rm rec}}{1100}\right)^{\frac{5}{2}}\left(\frac{\rm nG}{B_\lambda}\right)^2\left(\frac{h}{0.7}\right)^2\right]^{\frac{1}{n_{\rm B}+5}} \, {\rm Mpc}^{-1},\nonumber
\label{kDrec}
\end{eqnarray}
where $\omega_i\equiv\Omega_ih^2$ and $x_{\rm e}^{rec}$ is the fraction of free electrons at the time of recombination $z_{\rm rec}$. The justification for this expression is given in Appendix A. At any time after recombination, the evolution of the damping scale is presented as $k_{\rm D}(z)=k_{\rm D}^{rec}f_{\rm D}(z)$, where we obtain $f_{\rm D}(z)$  by integrating the energy conservation equation for PMFs:
\begin{equation}
-(1+z)H\frac{d\rho_{\rm mf}}{dz} = -4H\rho_{\rm mf} - \Gamma_{\rm mf},
\label{rho_mf}
\end{equation}
simultaneously with (\ref{Tbe})-(\ref{H}). In this equation, the first term on the right-hand side describes the adiabatic decay of the PMFs' energy density caused by expansion of the Universe ($\rho_{\rm mf}^{ad}\propto(1+z)^4$), and where $\Gamma_{\rm mf}=\Gamma_{\rm pmf}^{\rm (dt)}+\Gamma_{\rm pmf}^{\rm (amd)}$ is the dissipation rate of the PMFs' energy density in the post-recombination plasma due to the decaying magnetic turbulence and ambipolar diffusion. 
We set the initial condition for $f_{\rm D}(z_{rec})=1$. 

In such an approach, the magnetic field strength $B(z)$, smoothed at a scale $\lambda$ at any redshift $z$, is given as follows \citep{Minoda2019}:
\begin{equation}
 B^2_\lambda(z)=\frac{f_{\rm D}^{n_{\rm B}+3}(z)\cdot(1+z)^4}{2\pi^2}\int_0^\infty P_{\rm B}(k)e^{-k^2 \lambda^2}k^2 dk.
\end{equation}
We note $B^{ad}_{\lambda_{\rm D}}(z)\equiv B_0\cdot (1+z)^2$, where $\lambda_{\rm D}=2\pi/k_{\rm D}$. Then the energy density of the PMFs at any $z$ is 
\begin{eqnarray}
\rho_{\rm mf}(z)&=&f^{n_{\rm B}+3}_{\rm D}(z)\cdot\rho_{\rm mf}^{ad}(z)= \nonumber \\ 
&=&3.98\cdot10^{-20}\left(\frac{B_0}{\text{nG}}\right)^2f_{\rm D}^{n_{\rm B}+3}(z)\cdot(1+z)^4 \quad \frac{\rm erg}{\rm cm^3}.
\label{romf}
\end{eqnarray}
So, the factor $f_{\rm D}^{n_{\rm B}+3}$ describes the over-adiabatic decay of PMFs caused by heating the slightly ionised plasma after recombination. The often-used strength of smoothed PMFs  on the scale of $\lambda=1$ Mpc is related to $B_0$ as $B_{1\rm Mpc}=B_0(2\pi/k_{D})^{\frac{n_{\rm B}+3}{2}}$. So, for $\lambda_{\rm D}<1$ Mpc,  $B_{1\rm Mpc}<B_0$ and decreases with increasing $n_{\rm B}$.
 
\begin{figure}[htb]
\includegraphics[width=0.495\textwidth]{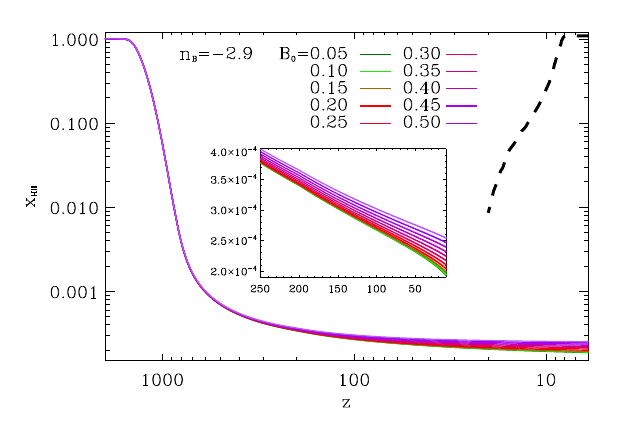}
\caption{Dependence of ionised hydrogen fraction on redshift in  models with primordial stochastic magnetic fields with different values of $B_0$, which increases for lines from bottom to top. The dashed line is the $2\sigma$ upper limit from the CMB radiation polarization data \citep{Planck2020a}.}
\label{xHII}
\end{figure}

\begin{figure*}[htb]
\includegraphics[width=0.495\textwidth]{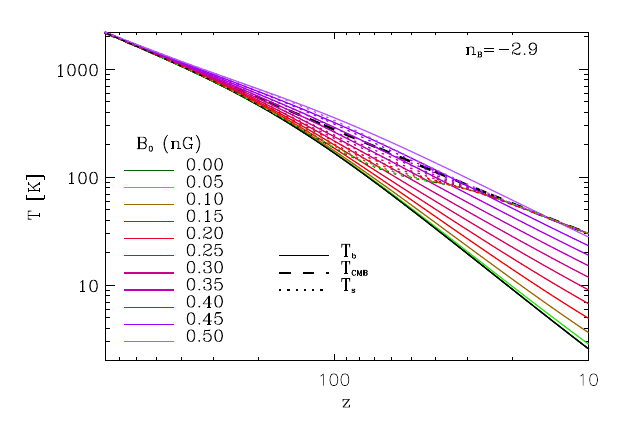}
\includegraphics[width=0.495\textwidth]{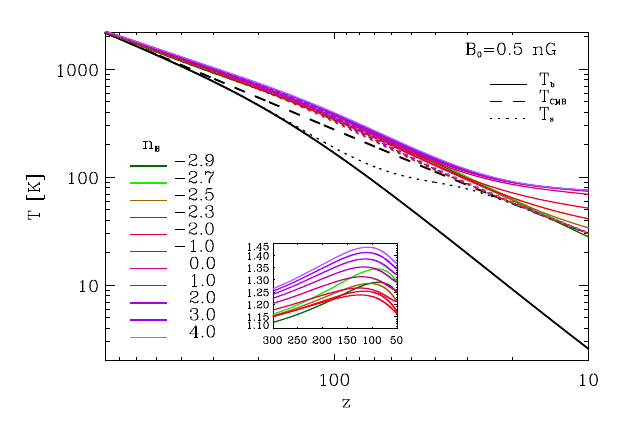}
\caption{Evolution of baryonic matter temperature $T_{\rm b}$ (solid lines) and spin temperature $T_{\rm s}$ (dotted lines) from cosmological recombination up to $z=10$ in $\Lambda$CDM model with PMF with different $B_0$ (left panel) and with different values of spectral indices $n_{\rm B}$ (right panel). The dashed line in both panels shows the temperature of CMB. The inset in the right panel shows $T_{\rm b}(z,B_0,n_{\rm B})/T_{\rm b}(z,B_0,n_{\rm B}=-2.9)$ for the purpose of distinguishing lines in the 21 cm line formation region.}
\label{Tb}
\end{figure*}

The decay of magnetic fields leads to the heating of baryonic matter, affecting the thermal and ionisation history of the Universe in the Dark Ages. Two mechanisms of heating of baryonic matter are considered: decaying magnetic turbulence and ambipolar diffusion. Taking into account the results in \cite{Mack2002,Sethi2005,Chluba2015,Kunze2014,Minoda2019}, the dependence of the corresponding heating functions on redshift can be represented as follows: 
\begin{eqnarray}
&&\Gamma_{\rm pmf}^{\rm (dt)}(z\ge z_{\rm rec}) = 1.5\rho_{\rm mf}^{ad}H(z)\frac{w_{\rm B}}{a}[f_{\rm D}(z)]^{n_{\rm B}+3}\times \nonumber \\
&&\hskip1.5cm\exp\left\{-\frac{(z-z_{\rm rec})^2}{5000}\right\}\left(\frac{1+z_{\rm rec}}{1+z}\right)^4, \label{Gmfdt1} \\
&&\Gamma_{\rm pmf}^{\rm (dt)}(z<z_{\rm rec}) = 1.5\rho_{\rm mf}^{ad}H(z)[f_{\rm D}(z)]^{n_{\rm B}+3}\times \nonumber \\
&&\hskip1.5cm\frac{w_{\rm B}a^{w_{\rm B}}}{(a+1.5\ln((1+z_{\rm rec})/(1+z)))^{w_{\rm B}+1}}, \label{Gmfdt2}\\ 
&&\Gamma_{\rm pmf}^{\rm (amd)}(z) = \frac{1-x_{\rm HII}}{g(T_{\rm b})x_{\rm HII}}[f_{\rm D}(z)]^{2n_{\rm B}+8}\times \nonumber \\
&&\hskip3.5cm\left[\frac{(1+z)k_{\rm D}^{rec}}{3.086\cdot10^{24}}\frac{\rho_{\rm mf}^{ad}}{\rho_{\rm b}}\right]^2f_{\rm L}, \label{Gmfad}
\end{eqnarray}
where $[\Gamma]$=erg/cm$^3$/s, $z_{\rm rec }= 1088$, and $t_{\rm rec}$ are the redshift and time of cosmological recombination accordingly;  $a = \ln(1+t_\mathrm{d}/t_\mathrm{rec})$, $w_{\rm B} \equiv 2(n_{\rm B}+3)/(n_{\rm B}+5)$, $t_\mathrm{d}/t_\mathrm{rec}=14.8/(B_0k_{\rm D})$,  and $\rho_{\rm b} = \rho_{cr}^{(0)}\Omega_{\rm b}(1+z)^3$ is the baryonic matter density; $g(T_{\rm b}) = 1.95\cdot10^{14} T_{\rm b}^{0.375}$~cm$^3$/s/g, and $f_{\rm L}=0.8313(n_{\rm B}+3)^{1.105}(1.0-0.0102(n_{\rm B}+3))$ is the analytic approximation of the Lorentz force integral \citep{Chluba2015}. 
 
With regard to the value of the spectral index $n_{\rm B}$, there are no strong theoretical constraints placed on it. In fact, the spectral index serves as the major discriminating factor among the generation mechanisms of primordial magnetic fields, given that different mechanisms inherently produce fields with different values of the spectral index. A variety of hypothetical generation mechanisms during the inflation epoch are considered in the literature; almost all of them introduce extra non-standard fields and/or interactions. Some of them include coupling of electromagnetic field to scalar fields, such as inflaton or dilaton \citep{Subramanian2009,Bamba2004, Durrer2023}; coupling to space-time curvature \citep{Kushwaha2020}; generation in Higgs-Starobinsky inflation \citep{Durrer2022}; coupling to axion-type fields \citep{Adshead2016}; introducing an extra-dimensional scale factor \citep{Kunze2005}; and considering spatial anisotropy in the background metric \citep{Pal2023}. There is also a possibility of magnetic field generation during phase transitions of the Universe. Scientists have done a lot of work to study this scenario, while the results for generated magnetic fields and spectral indices are strongly model dependent \citep{Vachaspati2021}. Finally, there are some models with primordial black holes that allow late field generation after primordial nucleosynthesis \citep{Saga2020,Papanikolaou2023}.
 
Nevertheless, we want to consider values of $n_{\rm B}$ that could be detectable with other observational methods. To be observable with current Planck constraints from CMB anisotropy, for example, the spectral index should not exceed the range $n_{\rm B} \simeq $ 1--2 \citep{Planck2016}. However, it should be noted that there is still a possibility that the true value of the spectral index is higher than this \citep{Durrer2003}. For this reason, we are interested in the range of $n_{\rm B} \in [-2.9,+4]$ and $B_0 \in [0.05, 0.5]$ nG. The expression for the heating function via ambipolar diffusion  (\ref{Gmfad}) is correct in the range of spectral index $-3\le n_{\rm B}\le5$ \citep{Kunze2014}.
 
In Fig. \ref{fD}, we present $f_{\rm D}(z)$ in power $n_{\rm B}+3$ for different $n_{\rm B}=-2.9$, ... 4. It depends on the amplitude of the magnetic field very weakly, therefore, we show it only for $B_0=0.5$ nG. The curves also illustrate the decrease in the PMFs' energy density caused by decaying magnetic turbulence and ambipolar diffusion over and above the adiabatic losses due to the expansion of the Universe: $f_{\rm D}(z)^{n_{\rm B}+3}=\rho_{\rm mf}(z)/\rho_{\rm mf}^{ad}(z)$. They decrease when $n_{\rm B}$ increases to $\sim-1$, then slightly increase to some approximant curve as the spectral index continues to grow. We explain this by the fact that the power spectrum $\propto k^{n_{\rm B}}$ with damping scale $k_D$ becomes similar to the delta-function when $n_{\rm B}\rightarrow +\infty$.

In Fig. \ref{Qpmf}, we show the dependencies of heating functions caused by decaying magnetic turbulence and ambipolar diffusion on redshift in models with different values of $B_0$ and $n_{\rm B}$. Both increase with increasing $B_0$. But the first one weakly depends on the value of $n_{\rm B}$, while the heating function caused by ambipolar diffusion quickly increases with increasing $n_{\rm B}$. Nevertheless, ambipolar diffusion at $z>100$ makes a negligible contribution to baryonic gas heating  compared to the decay of field turbulence $\Gamma_{\rm pmf}^{\rm (dt)}$, but it is comparable and large at lower $z$. In the inset of the right panel of Fig. \ref{Qpmf}, the ratios $\Gamma_{\rm pmf}^{\rm (dt)}(z,B_0,n_{\rm B})/\Gamma_{\rm pmf}^{\rm (dt)}(z,B_0,n_{\rm B}=-2.9)$ are presented to illustrate the change in the dependencies of $\Gamma_{\rm pmf}^{\rm (dt)}$ on $n_{\rm B}$ in the redshift range $300\le z\le50$, and the existence of an approximant curve for $\Gamma_{\rm pmf}^{\rm (dt)}$ for $n_{\rm B}>0$. We also note that an approximant curve exists for $\Gamma_{\rm pmf}^{\rm (amd)}$ as well (right panel). The apparent convergence of the curves in the left panel is illusory, due to the logarithmic scale.

In Fig. \ref{xHII}, we show the dependencies of the ionised fraction of hydrogen on redshift during the Dark Ages for PMF models with different values of $B_0$ in the range 0.05-0.5 nG and $n_{\rm B}=-2.9$. They differ only slightly from $x_{\rm HII}(z)$ in the standard $\Lambda$CDM since the heating of the gas in the PMF models with $B_0=0.5$ and $n_{\rm B}=-2.9$ does not exceed $1.3T_{\rm CMB}$ (see figs. below). Helium remains completely neutral, so $x_{\rm e}=x_{\rm HII}$.   

The results of integrations of the system of equations, (\ref{Tbe})-(\ref{H}) with (\ref{Gmfdt1})-(\ref{Gmfad}), are presented in Fig. \ref{Tb} for the models with different root mean square values of field strength $B_0=0.05-0.5$ nG and $n_{\rm B}=-2.9$ (left panel), and for different values of $n_{\rm B}=-2.9 ... 4$ and   $B_0=0.5$ nG (right panel). One can see that PMFs with $B_0\ge0.05$ nG noticeably increase the temperature of baryonic matter at $z<300$ in comparison with temperature in the null field model. If $B_0\gtrsim0.35-0.40$ nG and $n_{\rm B}=-2.9$, then the temperature $T_{\rm b}$ throughout the Dark Ages epoch is higher than the temperature of the CMB. Increasing $n_{\rm B}$ for the same values of $B_0$ and $z$ causes a non-monotonic increase of the temperature amplitude that we see in the inset of the right panel of Fig. \ref{Tb}, where the ratios $T_{\rm b}(z,B_0,n_{\rm B})/T_{\rm CMB}(z)$ are shown. This behaviour is caused by the dependence of $\Gamma_{\rm pmf}^{\rm (dt)}$ on $z$ shown in the inset of Fig. \ref{Qpmf}. The baryon temperature is an integral of $\Gamma_{\rm pmf}^{\rm (dt)}$ on $z$, which explains their different dependencies on $z$, but similar ones on $n_{\rm B}$.
\section{Global signal in 21 cm line: sensitivity to primordial magnetic field parameters} 

\begin{figure}[htb]
\includegraphics[width=0.495\textwidth]{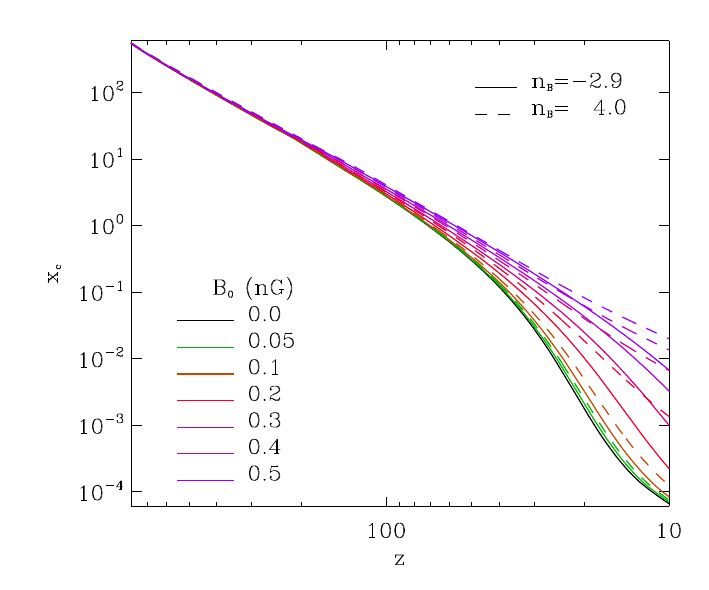}
\caption{Dependencies of collisional coupling $x_{\rm c}$ on redshift in the models $\Lambda$CDM+PMF with different values of $B_0$ and $n_{\rm B}$.}
\label{xc}
\end{figure}

\begin{figure*}[htb]
\includegraphics[width=0.495\textwidth]{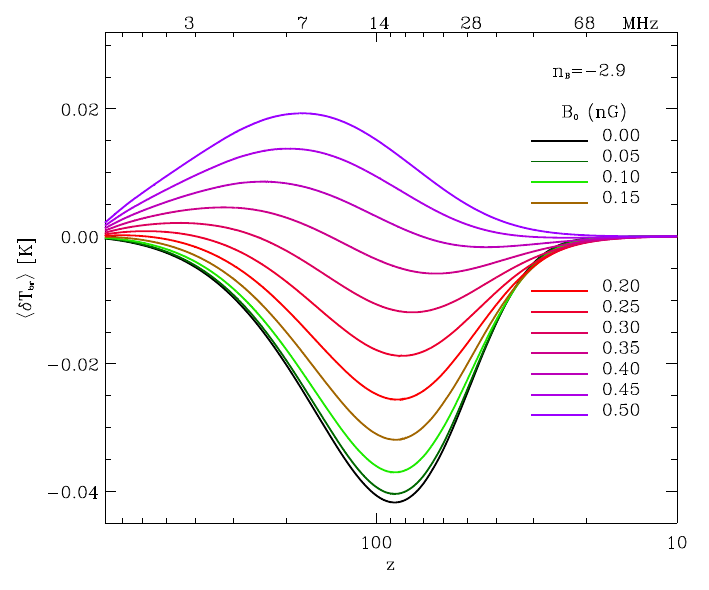}
\includegraphics[width=0.495\textwidth]{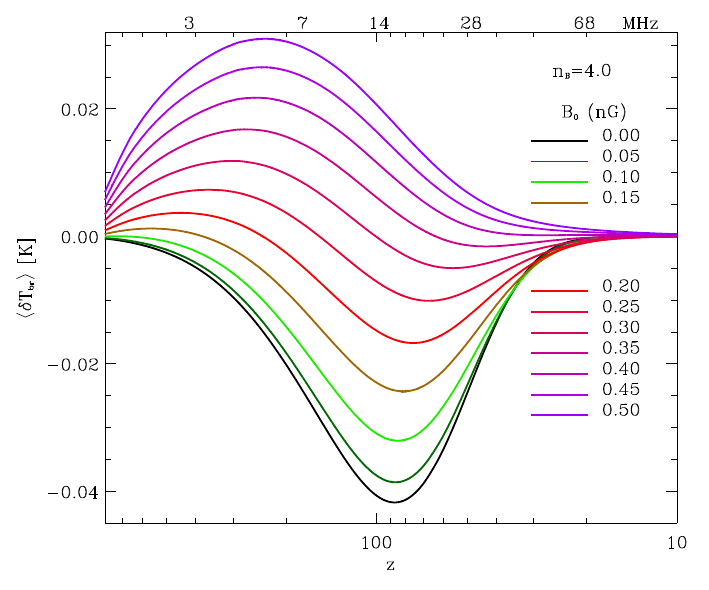}
\caption{Global signal in hydrogen 21 cm line for PMF models with $n_{\rm B}=-2.9$ (left panel) and $n_{\rm B}=4$ (right panel), with different values of $B_0$ (coloured lines).}
\label{dTbr_B}
\end{figure*}

\begin{figure*}[htb]
\includegraphics[width=0.495\textwidth]{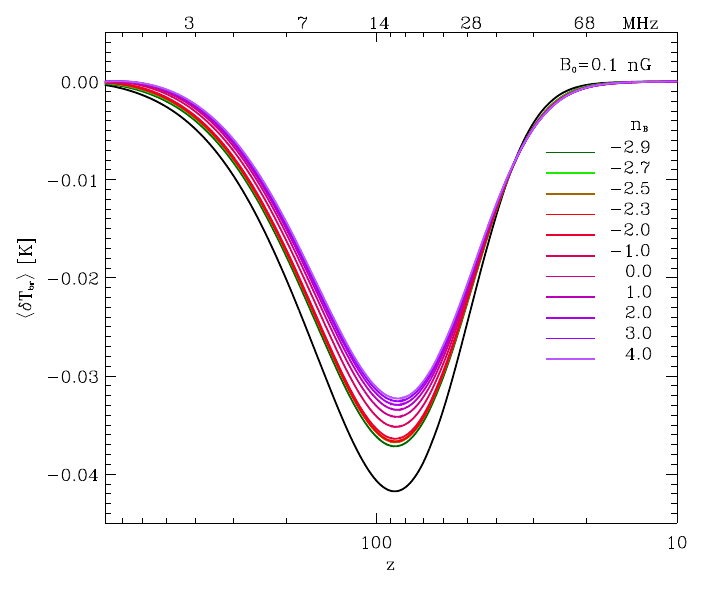}
\includegraphics[width=0.495\textwidth]{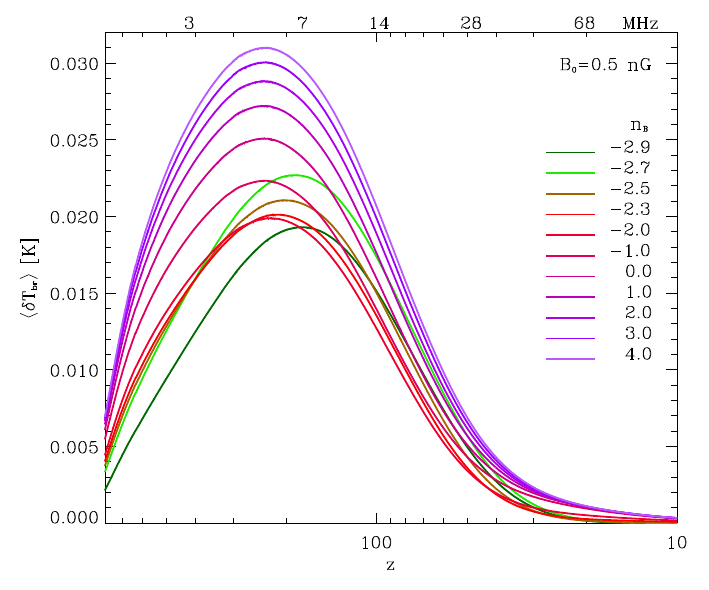}
\caption{Global signal in hydrogen 21 cm line for PMF models with $B_0=0.1$ nG (left panel) and $B_0=0.5$ nG (right panel), with different values of $n_{\rm B}$ (coloured lines).}
\label{dTbr_nb}
\end{figure*}

\begin{figure}[htb]
\includegraphics[width=0.495\textwidth]{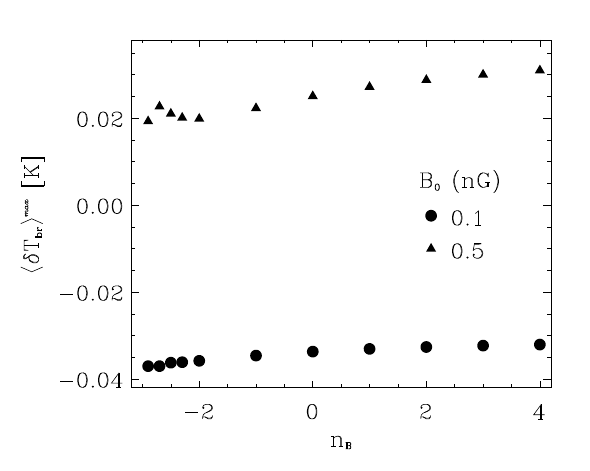} 
\caption{Dependence of amplitude of 21 cm line $\delta T_{\rm br}^{max}(n_{\rm B})$ on spectral index of PMF power spectrum for $B_0=$ 0.1 and 0.5 nG.}
\label{dTbr_nb_ampl}
\end{figure}
 
In this section, we analyse the dependence of the global signal in the 21 cm hydrogen line on PMF parameters. We have shown in previous papers \citep{Novosyadlyj2023,Novosyadlyj2024} that this line is noticeably sensitive to the heating of the intergalactic medium by PMFs when its amplitude is $\gtrsim0.1$ nG and $n_{\rm B}=-2.9$. For lower $B_0$, the global signal is an absorption line with differential brightness temperature $\delta T_{\rm br}\approx -41$ mK at frequency $\approx16$ MHz and the effective width of the line $\approx25$ MHz as in the standard $\Lambda$CDM model\footnote{In the papers \cite{Novosyadlyj2023,Novosyadlyj2024}, the amplitude of $\delta T_{\rm br}$ in the $\Lambda$CDM is cited as $\approx -35$ mK by number 23 instead of 27 in the expression for $\delta T_{\rm br}$.}. With increasing $B_0$, the depth of the absorption line decreases, disappears, and becomes an emission line at $B_0\approx0.3$ nG. In those papers, we supposed a nearly scale-invariant power spectrum of PMF with $n_{\rm B}=-2.9$. Since there are no observable constraints for the spectral index, it is sound to analyse the dependence of the brightness temperature in the line $\delta T_{\rm br}$ on $n_{\rm B}$ too.

Thus, we computed the evolution of the baryonic gas temperature $T_{\rm b}$ and the fraction of neutral hydrogen $x_{\rm HI}$ according to the prescription of Sect. 2: the excitation or spin temperature $T_{\rm s}$ and the differential brightness temperature $\delta T_{\rm br}$ as the solutions of the kinetic equations of excitation - de-excitation of hyperfine structure levels and radiation transfer equations accordingly,  
\begin{eqnarray}
&&\hskip-0.5cm\delta T_{\rm br}(z)=0.144 x_{\rm HI} \frac{\Omega_{\rm b}h}{\Omega_{\rm m}^{1/2}} (1+z)^{1/2} \left[1-\frac{T_{\rm CMB}}{T_{\rm s}}\right]\,\, \mbox{\rm K,}\quad \label{dTbr} \\
&&\hskip-0.5cm T_{\rm s}^{-1} = \frac{T_{\rm CMB}^{-1} + x_{\rm c} T_{\rm b}^{-1}}{1+x_{\rm c}}, \quad x_{\rm c} \equiv \frac{C_{10}}{A_{10}}\frac{h_{\rm P}\nu_{21}}{k_{\rm B} T_{\rm CMB}},\label{TsDA}
\end{eqnarray}
where $A_{10}$ is the Einstein spontaneous transition coefficient for the excited hyperfine structure level; $C_{10}$ is its collisional deactivation rate by electrons, protons, and neutral hydrogen atoms; $\nu_{21}=1420$ MHz is the rest frame frequency of the 21 cm line; $h_{\rm P}$ and $k_{\rm B}$ are Planck and Boltzmann constants accordingly. The total collisional deactivation rate $C_{10}$ depends on the temperature of the gas, the number densities of neutral hydrogen, protons, and electrons. It is crucial in the computation of the collisional coupling coefficient $x_{\rm c}$, which defines the spin temperature. The evolution of $x_{\rm c}$ in the $\Lambda$CDM model with PMFs is shown in Fig. \ref{xc}. It explains the evolution of the spin temperature $T_{\rm s}$ shown in Figs. \ref{Tb} by dotted lines. The relation between $T_{\rm s}$ and $T_{\rm CMB}$ at any $z$ defines the differential brightness temperature $\delta T_{\rm br}(z)$, which follows from eq. (\ref{dTbr}). When $T_{\rm s}\rightarrow T_{\rm CMB}$, then $\delta T_{\rm br}(z)\rightarrow0$. At $z>400$, the coupling coefficients are $x_{\rm c}\gg1$ and $T_{\rm s}\rightarrow T_{\rm b}$. But there $T_{\rm b}\approx T_{\rm CMB}$, so $\delta T_{\rm br}(z)\rightarrow0$ again.  

We computed the brightness temperature in the 21 cm hydrogen line using expressions (\ref{dTbr})-(\ref{TsDA}) for the same values of cosmological parameters and PMFs as in the previous section. The results are presented in Fig. \ref{dTbr_B} for different values of field strength $B_0=0.05$ ... 0.5 nG, and in Fig. \ref{dTbr_nb} for different values of spectral index $n_{\rm B}=-2.9$ ... 4.0. 

The results presented in Fig. \ref{dTbr_B} illustrate the dependence of the line profile on field strength, the transition from the absorption line to the emission line with increasing $B_0$. Fig. \ref{dTbr_nb} shows that the global signal in the 21 cm hydrogen line is also sensitive to the value of the spectral index $n_{\rm B}$. In Fig. \ref{dTbr_nb_ampl} we show the dependence of the amplitude of 21 cm line, $|\delta T_{\rm br}^{max}|$, on $n_{\rm B}$ for models with  $B_0=$ 0.1 and 0.5 nG. We note some non-monotonic dependence of spectral feature amplitude on $n_{\rm B}$ in the range $-2.9\le n_{\rm B}\le-2$ for models with $B_0=0.5$ nG.    
\begin{figure}[htb]
\includegraphics[width=0.495\textwidth]{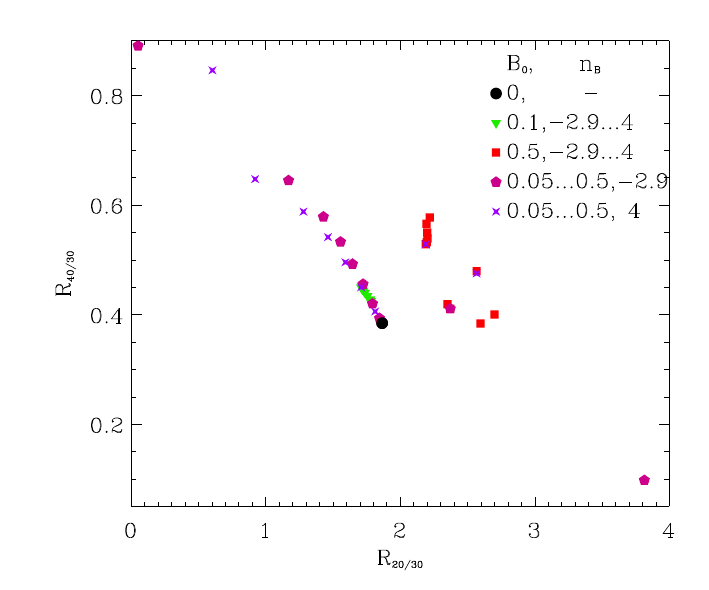}
\caption{Profiles of 21 cm lines in different ($B_0, n_{\rm B}$) models of PMFs in the space $R_{20/30}-R_{40/30}$.}
\label{R20/30}
\end{figure}

In the paper \cite{Okamatsu2024}, the authors propose using the high-frequency wing of the Dark Ages 21 cm line, especially the relation between two ratios  
\begin{eqnarray}
&&R_{20/30}\equiv\frac{\delta T_{\rm br}(\nu=20\,{\rm MHz})}{\delta T_{\rm br}(\nu=30\,{\rm MHz})} \quad {\rm and} \nonumber \\
&&R_{40/30}\equiv\frac{\delta T_{\rm br}(\nu=40\,{\rm MHz})}{\delta T_{\rm br}(\nu=30\,{\rm MHz})},\nonumber
\end{eqnarray}
as a measure of the deviations of the cosmological models from the  standard $\Lambda$CDM one and the possible distinction of  them. In Fig. \ref{R20/30} we present the relation $|R_{20/30}|-|R_{40/30}|$ for all profiles in Figs. \ref{dTbr_B}-\ref{dTbr_nb}. One can see that all $\Lambda$CDM+PMF models deviate from the $\Lambda$CDM one mainly along the diagonal $|R_{40/30}|^{max}-R_{20/30}|^{max}$, similar to other models in \cite{Okamatsu2024} (Fig. 4), but the spreads for the models analysed here are wider.

\section{Comparison and discussions}
Let us compare our results with similar ones recently published by other authors, in particular \cite{Minoda2019,Okamatsu2024,Mohapatra2025}. 
First, we note that the evolution of baryonic gas temperature and the 21 cm signal during the Dark Ages in the $\Lambda$CDM model with close parameters is practically the same as in these and other papers cited here. Comparison of results for PMFs models is not so unambiguous due to the authors' use of different approximations in calculations of decaying turbulence and ambipolar diffusion, the cut-off scale of the magnetic field at the time of recombination, taking into account its changes over time during the Dark Ages, and so on. In this paper, we completely describe the PMFs model and all the parameters and approximations used in the calculations. We even provide in Appendix A the derivation of the dependence of $k_{\rm D}$ on the cosmological parameters, since it is different from paper to paper.

In PMF models with $n_{\rm B}=-2.9$ and $B_0=0.1$, 0.2, 0.5 nG, Minoda et al. (2019)  obtained baryonic temperature $T_{\rm b}$ at $z=10$, approximately 4, 10, and 50 K (see Fig. 1 in \cite{Minoda2019}). Our values in the left panel of Fig. \ref{Tb} are ~4, 8, 37 K accordingly. The small differences are caused by the small differences in the parameters of the fiducial cosmological models and $k_{\rm D}^{rec}$. Mohapatra et al. (2025) obtained $T_{\rm b}$ with the same $k_{\rm D}^{rec}$ at z=10  $\sim$4, 10, and 30 K for PMF models with $n_{\rm B}=-2.99$ and $B_0=0.1$, 0.3, 0.55 nG respectively (see Fig. 3a in \cite{Mohapatra2025}). So, we can conclude that for an almost scale-invariant spectrum, the results are close. 

But if we compare our results for $n_{\rm B}=-2.5$ and $B_0=0.1$, 0.3 nG with those corresponding to \cite{Mohapatra2025} (Fig. 3a), then we see a strong disagreement despite the same expressions for $\Gamma_{\rm pmf}^{\rm (dt)}$ and $\Gamma_{\rm pmf}^{\rm (amd)}$. Their results show a steep increase in $T_{\rm b}$ with a small increase in $n_{\rm B}$, while our results show a slow increase in $T_{\rm b}$ (and $\delta T_{\rm br}$, Fig. \ref{dTbr_nb_ampl}) with $n_{\rm B}$. The authors in \cite{Mohapatra2025} did not explain such behaviour. This may be the case if the heating due to ambipolar diffusion is overestimated by several orders of magnitude, as can be assumed on the basis of Fig. \ref{Qpmf}.    

Comparing the profiles of the redshifted 21 cm line from the Dark Ages in the PMF models (Fig. \ref{dTbr_B}) with profiles in the models with self-annihilating and decaying dark matter particles (see Fig. 12 in \cite{Novosyadlyj2024}), one can see the similarities between them. It means that other signals from the Dark Ages must be used, together with the 21 cm hydrogen line redshifted to decametre wavelengths, to constrain the parameters of PMFs and decaying dark matter. These can be signals in the ro-vibrational lines of the first molecules \citep{Novosyadlyj2025,Kulinich2025}, the Cosmic Dawn 21 cm line redshifted to metre wavelengths \citep{Sethi2005,Tashiro2006,Schleicher2009,Kunze2014,Chluba2015,Kunze2019,Cruz2024,Bhaumik_2025}, Faraday rotation measurements \citep{Sullivan_2020}, magnetic fields' impact on CMB \citep{Paoletti2022}, Ly-$\alpha$ forest method \citep{Pavicevic2025} and so on.

Last but not least, the question is about the possibility of detecting the impact of PMFs on the 21 cm signal formed in the Dark Ages. It was mentioned in the Introduction that detection of the 21 cm signal from the Dark Ages needs about 1500 hours of integration time by ground-based Ukrainian radio telescope UTR-2 \citep{Konovalenko2024}. Nevertheless,  observations in the $\sim 10$–$50\,$MHz band are particularly challenging, as the 21 cm signal from the Dark Ages is strongly contaminated by several sources of foreground.

First, artificial, human-made sources, such as short-wave broadcasting, aviation communication, and amateur radio, produce radio frequency interferences that are many orders of magnitude stronger than the cosmological signal. Second, the ionosphere contributes a strong, time-dependent foreground that varies with the time of day. Third, our Galaxy emits synchrotron and bremsstrahlung radiation at these wavelengths, which constitutes a foreground with temperatures in order of $10^3-10^4$\,K. 

Recently proposed missions, such as DAPPER (Burns et al.
2021a) and FARSIDE (Burns et al. 2021b), designed to operate from the far side of the Moon, will be largely shielded from
the first two sources of contamination. In contrast, the Galactic synchrotron foreground, unlike the cosmological 21 cm signal, is polarised and exhibits a distinctive spatial structure, allowing for its effective removal through foreground subtraction techniques. Moreover, the spectral density of the galactic foreground is well described by the power-law frequency dependence \citep[see, e.g.,][]{Furlanetto2006}.

Interestingly, the non-detection of the 21 cm absorption line from the Dark Ages at $z\simeq100$ may allow us to put an upper limit on the  strength and spectral index of PMFs, if we overcome the degeneracy with alternative non-standard heating sources, for example, decaying or self-annihilating dark matter. We estimated that $\sim 3600$ hours of FARSIDE integration time would be sufficient to place an upper limit $0.1\,$nG on the magnetic field strength for $n_{\rm B} = -2.9$ or $n_{\rm B} = 4.0$ (see Appendix B for details). With 10~000 hours of integration time, we will be able to detect even the 21 cm absorption signal in the scenarios with $B_0\simeq0.2$\,nG. Moreover, within such an integration time it is possible to find signatures of the 21 cm emission for frequencies near 10 MHz ($z\simeq100$) in the scenarios with $B_0=0.5$\,nG and $n_{\rm B}=4.0$.

\section{Conclusions}
 We analysed the impact of primordial magnetic fields on the thermal history of the Universe in the Dark Ages due to the decaying magnetic turbulence and ambipolar diffusion in low-ionised plasma, and showed that fields with $B_0\ge0.05$ nG reduce the depth of the 21 cm absorption line of the standard $\Lambda$CDM model, in which the brightness temperature in line is $\delta T_{\rm br}\approx -40$ mK at frequency $\approx16$ MHz and the effective width $\approx25$ MHz. At $B_0\sim0.25-0.3$ nG, the absorption line becomes the emission line in the low-frequency wing of the absorption line. In models with $n_{\rm B}=-2.9$ and $0.05\le B_0\le0.5$ nG, the amplitude of the absorption/emission line ranges from $-40$ to $+20$ mK in the frequency range of $16-8$ MHz, the width at half-maximum $25-20$ MHz.
Thus, the global signal in the 21 cm line from the Dark Ages is sensitive to PMFs with a nearly scale-invariant power spectrum $n_{\rm B}=-2.9$ when $B_0\gtrsim0.05$ nG.

We found a sensitivity of the amplitude and profile shape of the 21 cm hydrogen spectral line formed during the Dark Ages to the value of the spectral index $n_{\rm B}$ at fixed values of the field strength $B _0$. As $n_{\rm B}$ increases from -2.9 to 4.0 in the case of PMFs with $B _0=0.1$ nG, the absorption line becomes shallower from -35 to -32 mK at the same frequency $\sim16$ MHz. In the case of PMFs with $B _0=0.5$ nG, the amplitude of the emission line increases from $\sim20$ to $\sim30$ mK. Comparisons with other gas-heating mechanisms in models outside the standard $\Lambda$CDM model revealed similarities in profiles with the decaying dark matter model, indicating the need to use other signals from the early Universe to resolve this degeneracy.
 
\begin{acknowledgements}
This work is carried out within the framework of the project \emph{“Tomography of the Dark Ages and Cosmic Dawn in the lines of hydrogen and the first molecules as a test of cosmological models”} (state registration number 0124U004029)
supported by the National Research Found of Ukraine.
\end{acknowledgements}

 \bibliographystyle{aa}
 \bibliography{refs.bib}

\begin{appendix}
\section{Damping scale of PMFs at the recombination epoch}

The damping scale of PMFs at the recombination epoch is defined in \cite{Subramania1998} as
\begin{equation}
\lambda_D=\sqrt{\frac{3}{5}}\frac{V_{\rm A}}{c}\lambda^\gamma_{\rm D},
\end{equation}
where the magnetohydrodynamic Alf{\' v}en velocity in units of light velocity at this epoch, according to \cite{Kunze2014}, equals  
\begin{equation}
 \frac{V_{\rm A}}{c}=3.24\cdot10^{-4}\left(\frac{B_\lambda}{\rm 1\, nG}\right)\left(\frac{k_{\rm D}}{k_\lambda}\right)^{\frac{n_{\rm B+3}}{2}}h^{-1},
\end{equation}
and baryon-photon Silk damping scale \citep{Silk1968,Peebles1970,Kaiser1983}:
\begin{equation}
\lambda^\gamma_{\rm D}=5.7\omega_m^{-\frac{1}{4}}\omega_b^{-\frac{1}{2}}\left(\frac{x_e^{rec}}{0.1}\right)^{-\frac{1}{2}}\left(\frac{1+z_{\rm rec}}{1100}\right)^{-\frac{5}{4}}\,{\rm Mpc}. 
\end{equation}
This approximation agrees well with the approximation proposed in \cite{Hu1997}. So,
\begin{eqnarray}
 \hskip-0.5cm&&\frac{1}{k_{\rm D}}\equiv\frac{\lambda_{\rm D}}{2\pi}=3.25\cdot10^{-4}\omega_m^{-\frac{1}{4}}\omega_b^{-\frac{1}{2}}\left(\frac{x_e^{rec}}{0.1}\right)^{-\frac{1}{2}}\nonumber\\
 \hskip-0.5cm&&\times\left(\frac{1+z_{\rm rec}}{1100}\right)^{-\frac{5}{4}}\left(\frac{B_\lambda}{\rm nG}\right)\left(\frac{k_{\rm D}}{k_\lambda}\right)^{\frac{n_{\rm B}+3}{2}}\frac{0.7}{h}\, {\rm Mpc}.
\end{eqnarray}
From this equation, we obtain the damping scale of PMFs (\ref{kDrec}) at the epoch of cosmological recombination.

\section{Sensitivity of lunar radio-telescopes to the impact of PMFs on 21 cm signal}\label{sec:appendix_obs}
We consider the FARSIDE mission \citep{Burns2021b}, proposed to operate for frequencies $\nu\leq40$ MHz, as a case study to evaluate the potential of constraining primordial magnetic field (PMF) parameters through observations of the 21 cm signal from the Dark Ages. The thermal noise of an ideal radiometer at frequency $\nu$ is given by \citep{Condon:2016}:
\begin{equation}
\sigma_\text{rms}(\nu) = \frac{T_\text{sys}(\nu)}{\sqrt{N \, \Delta\nu \, t}},
\end{equation}
where $T_\text{sys}(\nu)$ is the system temperature, approximately equal to the foreground temperature $T_\text{fg}$, $N$ is the number of antennas operating in autocorrelation mode, $\Delta\nu$ is the bandwidth, and $t$ is the integration time. The foreground emission can be modelled as a power law:
\[
T_\text{fg} = 5000\,\text{K}\left(\frac{\nu}{50\,\text{MHz}}\right)^{-2.5},
\]
following \cite{Burns2021b}; see also \citep{Furlanetto2006}. The FARSIDE mission concept includes 128 antennas with a bandwidth of $\Delta\nu = 0.5\,\text{MHz}$.

Detecting a $\sim 10^{-2}$\,K signal against a foreground with $T_\text{fg}\sim10^3$–$10^4$\,K is extremely challenging. A pragmatic strategy is to focus on identifying the extrema of the global absorption or emission profile (private communication with Y. Vasylkyvskyi, Institute of Radio Astronomy, NAS of Ukraine). Assuming a signal-to-noise ratio of $\text{SNR} \geq 10$, we estimate that detecting the 21 cm Dark Age signal in the absence of PMFs requires at least 1800\,hr of integration. With 5000\,hr integration time, FARSIDE could potentially detect the absorption signal predicted in a cosmological scenario with $n_{\rm B} = -2.9$ ($n_{\rm B} = 4.0$) and $B=0.2$\,nG ($B_0=0.15$\,nG). By contrast, detecting the 21 cm emission from $z\simeq200$ at $\nu\simeq 6$ MHz in the case of a non-helical magnetic field with $B_0=0.5$\,nG and $n_{\rm B}=-2.0$ (see Fig.~\ref{dTbr_nb}-\ref{dTbr_nb_ampl}) would require roughly $55\,000\,$hr, corresponding to more than six years of continuous observations.

In addition, for scenarios with $n_{\rm B} = 4.0$ and $B_0 = 0.3\,\mathrm{nG}$, where the signal exhibits two regimes, emission at $z \sim 200$, and absorption at $z \lesssim 100$, the FARSIDE array can detect the absorption signal with an integration time of $10{\,}000$\,hr and a signal-to-noise ratio of $\mathrm{SNR} = 10$. Notably, within the same observational strategy, this instrument can also detect emission at $z \lesssim 90$ for the scenario with $B_0 = 0.5\,\mathrm{nG}$ and $n_{\rm B} = 4.0$. 
\end{appendix}

\end{document}